\documentclass[twocolumn,showpacs,preprintnumbers,amsmath,amssymb,aps]{revtex4}

\usepackage{graphicx}
\usepackage{dcolumn}
\usepackage{bm}
\usepackage[caption=false]{subfig}
\usepackage{epsfig}
\usepackage{graphics}
\usepackage{graphicx}
\usepackage{mathtext}

\begin{document}

\title{Growing Scale-free Networks by a Mediation-Driven Attachment Rule}

\author{Md. Kamrul Hassan and Liana Islam}
\affiliation{University of Dhaka, Department of Physics, Theoretical Physics Group, Dhaka-1000,
Bangladesh.}

\begin{abstract}%

 We propose a model that generates a class of networks exhibiting power-law degree distribution with a
 spectrum of exponents depending on the number of links ($m$) with which incoming nodes join the existing
 network. Here, each new node first picks an existing node at random, and connects not with this but with
 $m$ of its neighbors also picked at random. Counter-intuitively enough, such a mediation-driven attachment
 rule gives rise to superhubs for small $m$ due to the {\it winners take all} effect, and hubs for large $m$
 due to the {\it winners take some} effect. To solve it analytically, we use mean-field approximation where
 we find that the inverse harmonic mean of degrees of the neighborhood of each existing node and their mean
 play a crucial role in determining the quality of the degree distribution.

\end{abstract}


\maketitle

In the recent past, we have amassed a bewildering amount of information on our universe, and yet we are
far from developing a holistic idea. This is because most of the natural and man-made real-world systems
are intricately wired and seemingly complex. Much of these complex systems can be mapped as networks
consisting of nodes connected by links, e.g., author collaboration and movie actor networks, the
Internet, World Wide Web, neural and protein networks etc. \cite{ref.coauthorship,ref.movieactor_2,
ref.internet, ref.www, ref.brain, ref.protein}. While working on some of these real-world
systems Barab\'{a}si and Albert (BA) in late $90$s found that the tail of the degree distribution
$P(k)$, the probability that a randomly chosen node is connected to $k$ other nodes, always follows a
power-law. In pursuit of theoretical explanation they realized that real networks are not static, rather
they grow. They also realized that a new node does not connect with an existing one randomly rather
preferentially with respect to their degrees - which is now known as the preferential attachment (PA)
rule \cite{ref.barabasi}. Incorporating both the ingredients BA proposed a model and showed that it
exhibits power-law degree distribution $P(k)\sim k^{-\gamma}$ where $\gamma=3$. Despite the profound
success of the BA model we are compelled to note a couple of drawbacks. First and foremost, the
preferential attachment (PA) rule is too direct. Second, the BA model can only account for $\gamma=3$
while in real networks the exponent $\gamma$ assumes a spectrum of values between $2<\gamma\leq 3$. To
overcome these drawbacks there exist a few variants of the BA model through the inclusion of rewiring,
aging, ranking etc. \cite{ref.rewiring, ref.rewiring_1, ref.aging_1, ref.ranking}, and a few
alternatives to the direct PA mechanism like redirection, vertex copying, and duplication mechanisms
\cite{ref.redirection_sublinear, ref.hi_dispersed, ref.copying, ref.duplication}.

\begin{figure}
\centering
\subfloat[]
{
\includegraphics[height=1.4in, width=1.4in, trim = 42mm 42mm 32mm 32mm, clip=true]{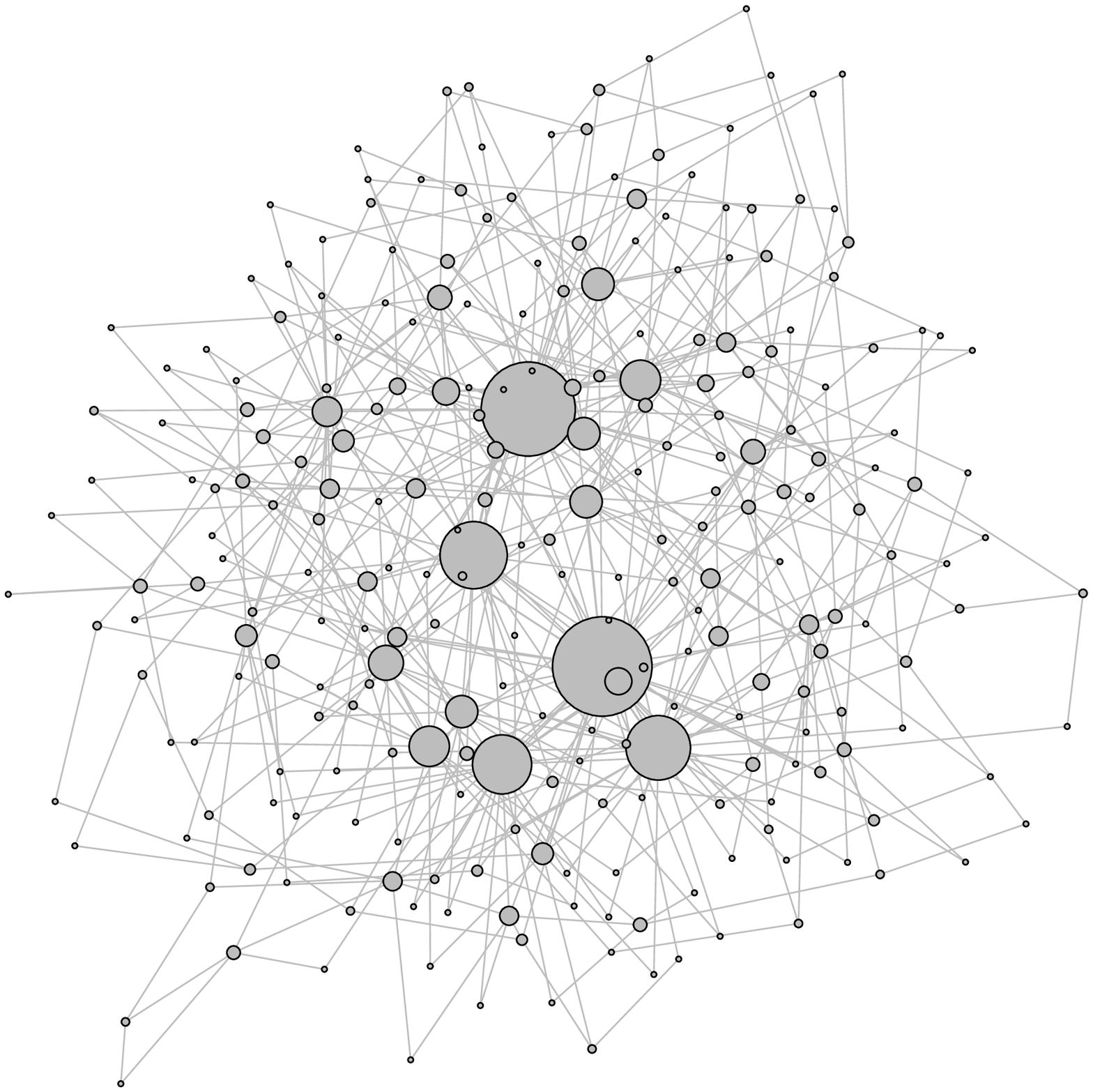}
\label{fig:a}
}
\subfloat[]
{
\includegraphics[height=1.4in, width=1.4in, trim = 42mm 42mm 32mm 32mm, clip=true]{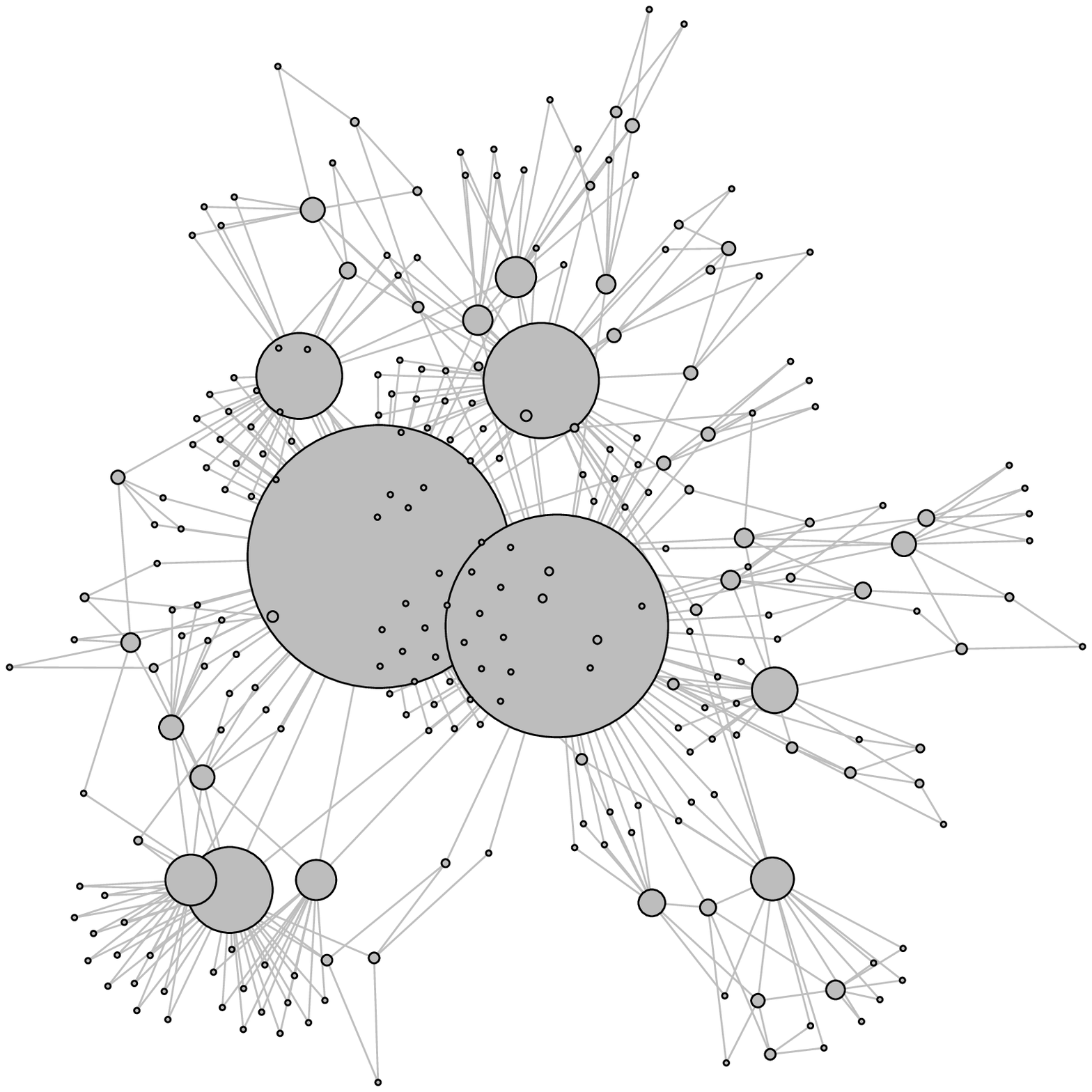}
\label{fig:b}
}
\caption{Snapshots of networks grown using (a) the PA rule of the BA model and (b) the MDA rule. The
size of each node is drawn according to its degree showing that the hubs of networks grown using the MDA
rule are richer than those of the BA networks.} \label{fig:ab}
\end{figure}

In this article, we propose a model in which a walker is parachuted to an arbitrary node of the existing
network at random and then takes a random step to one of its neighbors. 
The new node is attached to the node where the walker stops. The model has been
inspired by the growth of the weighted planar stochastic lattice (WPSL) \cite{ref.hassan_njp,
ref.hassan_conf}. Seeing its dual emerges as a scale-free network, we became curious as to what happens
when a graph is grown following a similar rule. That is, an existing node connects with the
incoming node only if one of its neighbors is picked at random. We call it the mediation-driven
attachment (MDA) rule since the node where the walker first landed acts as a mediator for connection
between its neighbor and the new node. We use mean-field approximation (MFA) to solve the model
analytically, and use numerical simulation to verify the solution. We show that for small $m$, the MDA
rule is super-preferential due to {\it winners take all} (WTA) effect that gives rise to a few
superhubs. The snapshots of the networks grown according to BA and MDA rules shown in FIG. \ref{fig:ab}
for $m=2$ clearly reveal the presence of hubs and superhubs respectively. However, for large $m$ the WTA
effect is replaced by {\it winners take some} (WTS) effect that gives rise to simple hubs only. The
essential idea of the MDA rule can be seen in the formation of trade links among businessmen and in the
growth of the WWW. In business, a newcomer wishing to establish trade links with other businessmen can
never assess the whole network to see who has how many links in order to decide who will be his partner,
owing to the large size of the trade world. Instead, he uses a mediator to find a suitable partner. We
note that there exist a couple of works similar to our model albeit the ensuing analysis and the results
are completely different from ours \cite{ref.mda_1, ref.mda_2, ref.redirection}.

\begin{figure}
\label{fig1}
\includegraphics[width=3.2cm,height=3.8cm, trim = 18mm 65mm 20mm 20mm, clip=true]{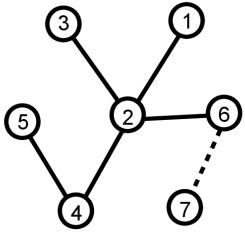}
\caption{A schematic description of the MDA rule.}
\label{fig2}
\end{figure}

To illustrate the MDA rule we consider a miniature network in FIG. \ref{fig2} consisting of $m_0=6$
nodes labeled $i=1,2,...,6$. This can be considered as the seed. Now to connect a new node to it, we
first pick a node at random from the $6$ existing nodes, say node $2$. Second, we pick one of its four
neighbors, also with {\it equal a priori} probability, say it is node $6$, and connect it with the new
node which we label $7$. In this particular example we assume that the new nodes are born with one link
or edge, i.e. $m = 1$. In the case of new nodes arriving with more than one edge we first pick one node
randomly from the entire network followed by picking $m$ of its distinct neighbors and connect them with 
$m$ new edges. Now the question is, according to our model, what is the probability $\Pi(i)$ that an
existing node $i$ is finally picked and the new node gets connected with it? Say, the node $i$ has
degree $k_i$ and its $k_i$ neighbors, labeled $1,2,\hdots ,k_i$, have degrees $k_1,k_2,...,k_{k_i}$
respectively. We can reach the node $i$ from each of these $k_i$ nodes with probabilities inverse of
their respective degrees, and each of the $k_i$ nodes can be picked at random with probability $1/N$. We
can therefore write
\begin{equation} \label{eq:1}
\Pi(i)= \frac{1}{N} \Big [ \frac{1}{k_1}+ \frac{1}{k_2} + \hdots + \frac{1}{k_{k_i}} \Big ] =
\frac{\sum_{j=1}^{k_i}{\frac{1}{k_j}}}{N}.
\end{equation}
The probabilities $\Pi(i)$ are normalized, which implies that
$\sum_{i=1}^N\sum_{j=1}^{k_i}{\frac{1}{k_j}}=N$.
We have verified it numerically and found that it is independent of $m$ and $N$. The rate at which an
arbitrary node $i$ gains links is therefore given by the following rate equation.
\begin{equation} \label{eq:2}
{\frac{\partial k_i}{\partial t}} = m \ \Pi(i)
\end{equation}
The factor $m$ takes care of the fact that any of the $m$ links of the newcomer may connect with the
node $i$.

\begin{figure}[htb]
\begin{center}
\includegraphics[width=5.0cm,height=8.5cm,clip=true,angle=-90]{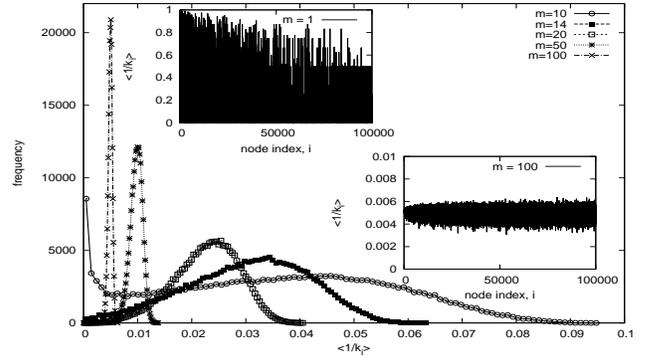}
\caption{The distribution of IHM for different $m$ values. The fluctuations of IHM of individual
nodes is shown in the inset (top left for $m=1$ and bottom right for $m=100$).}
\label{fig:histogram_inset_fluctuations}
\end{center}
\end{figure}

Solving Eq.\eqref{eq:2} for $\Pi(i)$ given by Eq.\eqref{eq:1} seems quite a formidable task unless we
can simplify it in some way. A few steps of hand calculation on a network of small size reveals that the
probability $\Pi(i)$ of picking node $i$ is higher for nodes with higher degree than those with lower
degree, encouraging us to re-write Eq.\eqref{eq:1} as
\begin{equation} \label{eq:npi}
\ \Pi(i) = \frac{k_i}{N} \ \frac{\sum_{j=1}^{k_i}{\frac{1}{k_j}}}{k_i}.
\end{equation}
The factor $\frac{\sum_{j=1}^{k_i}{\frac{1}{k_j}}}{k_i}$ is the inverse of the harmonic mean (IHM) of
degrees of the $k_i$ neighbors of a node $i$ which we denote as $\Big <\frac{1}{k_i}\Big >$ for
convenience. We now attempt to replace the IHM value of each node by an effective ``mean value''
\begin{equation}
\Big <{\frac{1}{k_i}}\Big >\approx \overline{\Big <{\frac{1}{k_i}}\Big >},
\end{equation}
where the overline indicates the ensemble average of the mean IHM of each realization. In this way, all
the information on correlations in the fluctuations is lost and hence we call it mean-field
approximation (MFA). One immediate consequence is that like the BA model, our MDA rule too is
preferential in character. This may sound a bit too drastic at this stage, but we shall justify it
anyway. It is needless to mention that this approximation will work only if the fluctuations of IHM of
individual nodes from their mean is not wild.

\begin{figure}[htb]
\begin{center}
\includegraphics[width=4.5cm,height=8.0cm,clip=true,angle=-90]{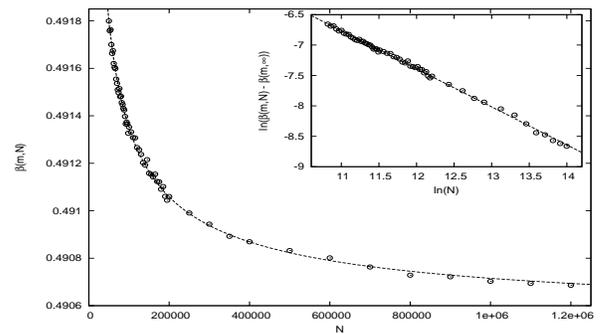}
\caption{Plot of $m\overline{\Big <{\frac{1}{k_i}}\Big >}\equiv \beta(m,N)$ vs. $N$ for $m=70$. Inset shows
that $\beta(m,N)$ saturates to $\beta(m,\infty) = 0.490513$ algebraically as $N^{-3/5}$.}
\label{fig:m70_beta_approximate_vs_N_verification_of_law}
\end{center}
\end{figure}

We perform extensive numerical simulation to check how robust the approximation is. For that we
calculate the IHM for each node of the network grown with fixed $m$, and plot them as a function of node
label $i=1,2,...,N$ (as in the insets of FIG. \ref{fig:histogram_inset_fluctuations}). We find that for
small values of $m$, the data points fluctuate so wildly from node to node that the mean of
$\Big <{\frac{1}{k_i}}\Big >$ over the entire size of the network bears no meaning, which can also be
seen from the frequency distributions of IHM shown in FIG. \ref{fig:histogram_inset_fluctuations}. The
frequency distributions become symmetric as we increase $m$ approximately beyond $14$, and the fluctuations occur at
increasingly lesser extents, and hence the meaning of the mean seems to be more meaningful. We also find
that the mean of the IHM depends on $m$ and is independent of network size $N$ in the limit
$N \rightarrow \infty$
(as demonstrated in FIG. \ref{fig:m70_beta_approximate_vs_N_verification_of_law}). So, we can write for
$m > 14$,
\begin{equation} \label{eq:approx}
\Pi(i) \approx \frac{k_i}{N} \ \overline{\Big <{\frac{1}{k_i}}\Big >} = \frac{k_i}{N} \ \frac{\beta(m)}{m},
\end{equation}
where the factor $m$ in the denominator is introduced for future convenience. Thus the value of
$\beta(m)$ for each fixed $m$ is also independent of $N$ as $N \rightarrow \infty$. Eq.\eqref{eq:approx}
confirms that the attachment probability $\Pi(i) \propto k_i$, and hence like the BA model our model too
is preferential in character, but not directly.

\begin{figure}[htb]
\begin{center}
\includegraphics[width=4.5cm,height=8.0cm,clip=true,angle=-90]{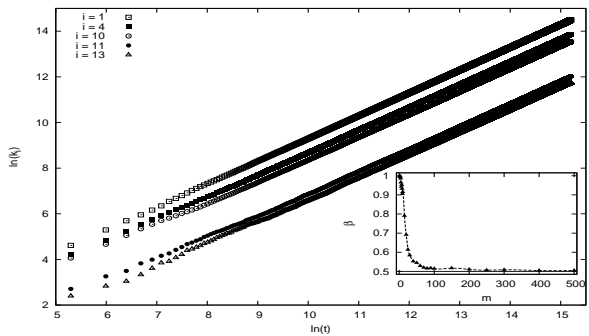}
\caption{Plots of $\ln(k_i)$ versus $\ln(t)$ for $5$ nodes added at five different times for $m = 2$. In
the inset, we show variation of the slope $\beta$ with $m$.}
\label{fig:kth2.eps}
\end{center}
\end{figure}

It is noteworthy that the size $N$ of the network is an indicative of time $t$ since we assume that only
one node joins the network at each time step. Thus, for $N>>m_0$ we can write $N \sim t$. We can now
solve the rate equation. Using  Eq.\eqref{eq:approx} and $N \sim t$ in Eq.\eqref{eq:2} we find that the
rate equation we ought to solve is
\begin{equation} \label{eq:3}
{\frac{\partial k_i}{\partial t}} = k_i \ \frac{\beta (m)}{t}.
\end{equation}
\noindent Solving it subject to the initial condition that the $i$th node is born at time $t=t_i$ with
$k_i(t_i)=m$ we get,
\begin{equation} \label{eq:4}
 k_i(t) = m \Big ({\frac{t}{t_i}}\Big )^{\beta(m)}.
\end{equation}
The form of the solution is exactly the same as that of the BA model \cite{ref.barabasi} except for the
$\beta$ value. To test Eq.\eqref{eq:4} we plot $\ln(k_i)$ versus $ln(t)$ in FIG. \ref{fig:kth2.eps} for
a few randomly selected nodes and find a set of parallel straight lines. It implies that all nodes grow
linearly as predicted by Eq.\eqref{eq:4}. However, in contrast to the BA model, the slope ($\beta$) of
the straight lines depends on $m$, asymptotically approaching $0.5$ for large $m$ (see inset of FIG.
\ref{fig:kth2.eps}). Also, there is a significant difference between $\beta$ obtained using our
approximation $\beta(m)= m\overline{\Big <{\frac{1}{k_i}}\Big >}$, with that obtained from the
$\ln(k_i)$ vs. $\ln(t)$ plots when $m\le14$. The difference gets negligible as $m$ is increased and this
too is correlated to the increase in the symmetry and decrease in the width of the IHM distributions.

\begin{figure}[htb]
\begin{center}
\includegraphics[width=5.0cm,height=8.5cm,clip=true,angle=-90]{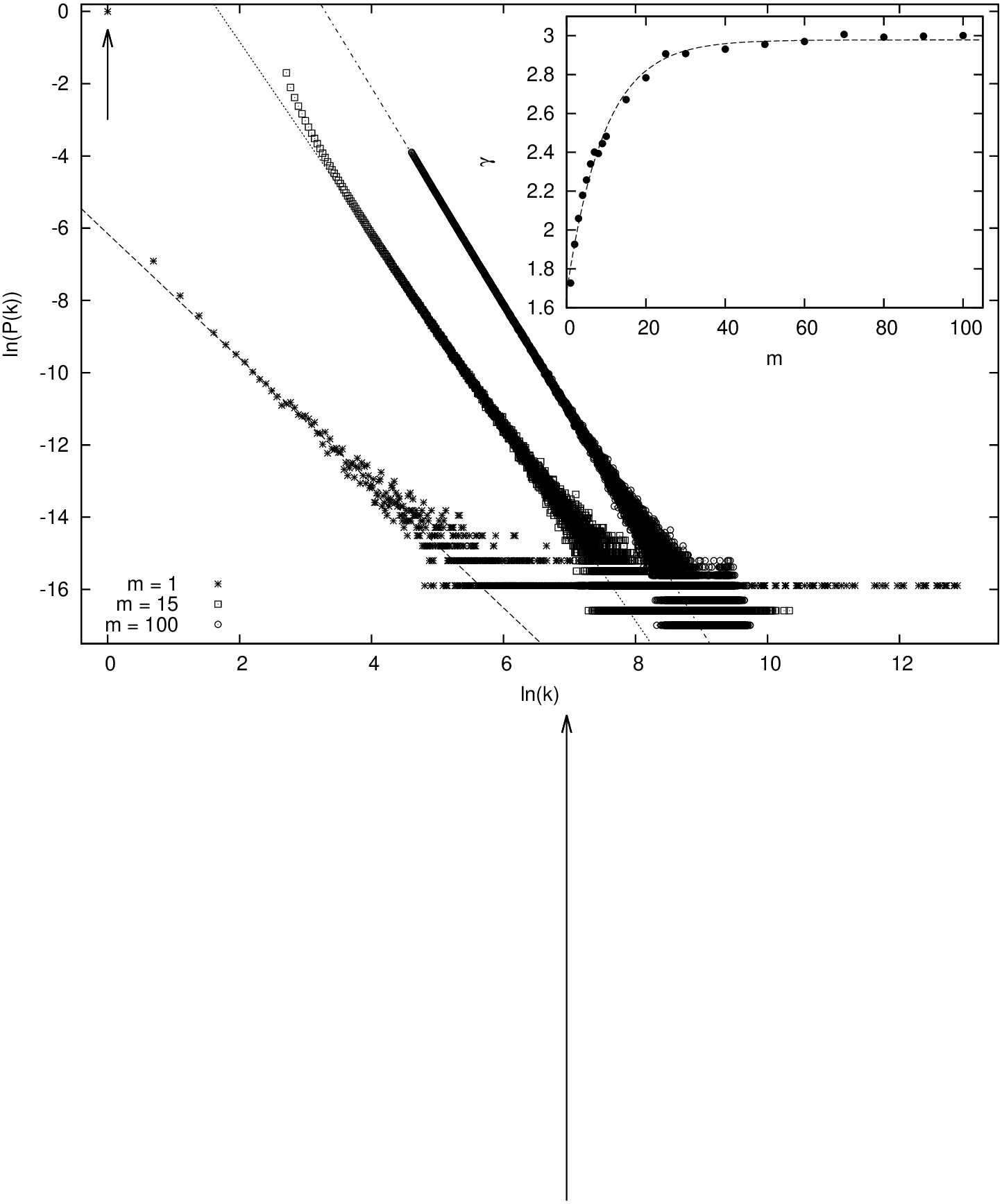}
\caption{Plots of $\ln (P(k))$ vs $\ln(k)$ for $m = 1$, $m=15$ and $m=100$ revealing that the MDA rule
gives rise to power-law degree distribution. In the inset we show the variation in the exponent
$\gamma$ as a function of $m$.}
\label{m15m100lnprE40.eps}
\end{center}
\end{figure}

We can now find the degree distribution $P(k)$ by appreciating the fact that it is related to $P(t_i)$,
the probability with which node $i$ is added to the system at time $t_i$, via
\begin{equation} \label{eq9}
P(k) \ dk=-P(t_i) \ dt_i.
\end{equation} 
The minus sign is introduced here to take into account that the smaller the value of $t_i$ the larger
the degree $k_i$. Substituting $P(t_i) =\frac{1}{N} \sim \frac{1}{t}$ (since new nodes are added to the
system at equal interval of time) and the derivative of $t_i$ with respect to $k_i$ from Eq.\eqref{eq:4}
in Eq.\eqref{eq9} gives
\begin{equation} \label{eq:20}
 P(k) \sim k^{-\gamma(m)}, \: \text{where} \: \: \gamma(m) = \frac{1}{\beta(m)}+1.
\end{equation}
The most immediate difference of this result from that of the BA model is that the exponent $\gamma$
depends on $m$. To verify it we plot in FIG. \ref{m15m100lnprE40.eps} $\ln(P(k))$ vs. $\ln(k)$ using
data extracted from numerical simulation and find straight lines with characteristic fat-tail. This
confirms that our analytical solution is in agreement with the numerical solution. It is important to
note that for small $m$, especially for $m=1$ there is a special point in the plot of the degree
distribution that is way far above the other points. This special point corresponds to $99.5\%$ of the
all the nodes which have degree $1$, and which are held by a few hubs. This is reminiscent of the WTA
phenomenon. However, as $m$ increases, we find that the WTA phenomenon is replaced by the WTS phenomenon
and at the same time, all the data points follow the same trend. This happens when the IHM has less
noise and the mean of IHM has meaning. Therefore, we argue that the IHM can be regarded as a measure of
how curved or straight the degree distribution is.

The next thing we do is to give a generalized analytical expression for $\gamma$. We fit the data and
find that $\gamma$ saturates to $3$ as $\gamma = 3(1 - 0.43531 e^{-0.11m})$. The value of $m$ for which
$\gamma$ has increased to within a factor of $1-e^{-1} (\approx63 \%)$ of its maximum value $3$, is
$m_{c} = \frac{1}{0.11} \approx 9$. We can relate this to the frequency distribution of the IHM in a way
that below $m=9$ the distribution is highly asymmetric. From $m=8$ to $m=14$, the distributions
gradually metamorphose to assume a Gaussian shape. Also, our numerical simulations show that the
whole $\gamma$ spectrum for $m\ge3$ is between $2$ and $3$. Our result stands in sharp contrast with
that of Yang et al. who found the exponent $\gamma$ to vary in the range $1.90$ to $2.61$ when $m$ was
varied in the range $1$ to $200$ \cite{ref.mda_1}. Besides, our expression for the probability $\Pi$ and
method of solution are very different from theirs.

To summarize, we have proposed a new attachment rule, namely the MDA rule, for growing networks that
exhibit power-law degree distribution. At a glance, it may seem MDA defies the intuitive idea of the PA 
rule, however, a closer look reveals otherwise.
Indeed, we show explicitly that the MDA rule is in fact not only preferential but superpreferential for $m<14$
and for alrge $m$ it embodies the PA rule but in disguise. We obtained an exact expression for
$\Pi(i)$ that describes the probability with which an existing node $i$ is finally picked to be
connected with a new node. Solving the model analytically for the exact expression of $\Pi(i)$ appeared
to be a formidable task. However, the good news is that we could still find a way to use MFA. Later it
turns out that not being able to solve analytically for the exact expression for $\Pi(i)$ was highly
rewarding as it helped us gain a deeper insight into the problem. While working with the expression for
$\Pi(i)$, we realized that the IHM value of the degrees of the neighborhood of node $i$ plays a crucial
role in the model. We find that the fluctuations in IHM values of existing nodes are so wild for small
$m$ that the mean, in this case, bears no meaning and hence MFA is not valid in this case. We observed that
for small $m$ such as $m=1$ or $2$, the MDA rule becomes super-preferential in the sense that more than
$97\%$ of the nodes are extremely poor in degree as they are linked to one or two other nodes which are
super-hubs. This is reminiscent of the WTA phenomenon. However, for large $m$, the fluctuations get
weaker and their distribution starts to peak around the mean revealing that the mean has a meaning. This
is the regime where MFA works well and we verified it numerically. Here we found that the WTA phenomenon
is replaced by the WTS phenomenon. We hope to extend our work to study dynamic scaling and universality
classes for the MDA model for different $m$ and check how it differs from the BA model
\cite{ref.mda_data_collapse}.

\hfill \break

We acknowledge Syed Arefinul Haque for his technical support.

\end{document}